\documentclass[12pt]{article}
\usepackage{latexsym}
\usepackage{amsmath,amsbsy,amssymb}

\newcommand{\dd}{\mbox{\rm d}}
\newcommand{\wg}{\wedge}
\newcommand{\gam}{\gamma}
\newcommand{\Gam}{\Gamma}
\newcommand{\dg}{\dagger}
\newcommand{\ddg}{\ddagger}
\newcommand{\tl}{\tilde}
\newcommand{\ul}{\underline}
\newcommand{\nl}{\natural}

\newcommand{\bth}{\bar \theta}
\newcommand{\nth}{\theta}

\newcommand{\DD}{\mbox{\rm D}}

\newcommand{\p}{\partial}

\newcommand{\be}{\begin{equation}}
\newcommand{\bear}{\begin{eqnarray}}
\newcommand{\ear}{\end{eqnarray}}
\newcommand{\ee}{\end{equation}}
\newcommand{\lbl}{\label}
\newcommand{\bi}{\bibitem}
\newcommand{\ci}{\cite}

\newcommand{\vs}{\vspace}

\begin{document}

\

\baselineskip .7cm 

\vs{15mm}

\begin{center}

{\LARGE \bf  Space Inversion of Spinors Revisited: A Possible
Explanation of Chiral Behavior in Weak Interactions
}

\vs{3mm}

Matej Pav\v si\v c

Jo\v zef Stefan Institute, Jamova 39,
1000 Ljubljana, Slovenia

e-mail: matej.pavsic@ijs.si

\vs{6mm}

{\bf Abstract}
\end{center}

\baselineskip .5cm 

{\small

%\vs{.1cm}

We investigate a model in which spinors are considered as being
embedded within the Clifford algebra that operates on them.
In Minkowski space $M_{1,3}$, we have four independent
4-component spinors, each living in a different minimal
left ideal of $Cl(1,3)$. We show that under space inversion,
a spinor of one left ideal transforms into a spinor of
another left ideal. This brings novel insight to the
role of chirality in weak interactions. We demonstrate the latter role 
by considering an action for a generalized spinor field
$\psi^{\alpha i}$ that has not only a spinor index $\alpha$
but also an extra index $i$ running over four ideals.
The covariant derivative of $\psi^{\alpha i}$ contains the
generalized spin connection, the extra components of which are
interpreted as the SU(2) gauge fields of weak interactions and their
generalization.
We thus arrive at a system that is left-right symmetric
due to the presence of a ``parallel sector", postulated
a long time ago, that contains mirror particles coupled
to mirror SU(2) gauge fields.
}



\normalsize 

\baselineskip .59cm 

\section{Introduction}

After the seminal paper by Lee and Yang\,\ci{LeeYang}, we have become
accustomed to the idea that the physical processes associated
with weak interactions are not invariant under space inversion.
However, in this same paper, Lee and Yang proposed that the introduction of
mirror particles restores the invariance of $\beta$-decay under
space inversion. The idea of mirror particles and the exact parity
model has been thereafter pursued and investigated
from various points of view in Refs.\,\ci{Kobzarev}--\ci{Foot}.
The possibility that mirror particles
are responsible for dark matter has been explored in many
works\,\ci{Dark}. For a review on mirror particles and their phenomenological
consequences, see \ci{OkunRev}.

In this paper, we will show how mirror particles find a natural
explanation if we employ the concept of algebraic spinors.
It is well known that spinors can be described as elements of the left or right
minimal ideals of a Clifford algebra\,\ci{Riesz, Teitler}.
As observed in Ref.\,\ci{Holland},
such an approach yields a more general and coherent theory than the
traditional ``column" representation. It incorporates the usual
spinor formalism, but in addition, it brings  the features that
are absent from the usual theory in which spin space is a representation
space lying outside of the algebra that operates on it. In a more
general theory, a spinor can be considered as being embedded within
the Clifford algebra that operates on it. This opens a Pandora's box
of possibilities that have been explored in the attempts to find a
unified theory of fundamental particles and
forces\,\ci{CliffUnific}--\ci{PavsicE8}. 

We will follow the approach\,\ci{SpinorNilpot,Winnberg} in which
spinors are constructed in terms of nilpotents formed from the spacetime
basis vectors represented as generators of the Clifford algebra $Cl(1,3)$.
We find that under space inversion, a left-handed spinor of one
ideal  transforms into a right-handed spinor of not
the same but of a different ideal. We then construct a specific
model starting from a generalized Dirac action that is a functional
of 16 complex valued fields, components of a generalized
algebraic spinor field. Such a 16-component field is written as
the $4 \times 4$ matrix $\psi^{\alpha i}$, representing four Dirac spinors lying
in four different left ideals. A transformation R that is generated
by bivectors $\gamma_\mu \wg \gam_\nu$, $\mu,\nu=0,1,2,3$, can act on 
$\psi^{\alpha i}$ from the left, in which case R has the role of
a Lorentz transformation, or from the right, in which case it behaves
as an `internal' transformation. Our model thus incorporates the
internal gauge group SU(2). The covariant derivative occurring in the
action contains a generalized spin connection that incorporates
not only the ordinary spin connection but also two kinds of
SU(2) gauge fields, $W_\mu$ and $W_\mu^\nl$, that are transformed
into each other by the space inversion P. The `internal' components
of $\psi^{\alpha i}$, denoted by the index $i$ that runs over four
different left ideals, form two irreducible representations
of SU(2) that correspond to ordinary and mirror particles.
Thus, we have found a theoretical explanation for the suggestion that there
must be a mirror world---a parallel sector---with no common
interaction with ordinary particles\,\ci{Kobzarev,Foot}.
In this paper, we demonstrate the existence of such a parallel sector
for the case of weak interactions only.

\section{Clifford algebra and spinors in Minkowski space}

In constructing spinors, we will follow Ref.\,\ci{Winnberg}.
Let us consider the Minkowski space $M_{1,3}$ in which a quadratic
form is $x^\mu \eta_{\mu \nu} x^\nu = x^2$, where $\eta_{\mu \nu} =
{\rm diag} (+ - - -)$ is the Minkowski metric tensor. The square root
of $x^2$ is $x= x^\mu \gam_\mu$, where $\gam_\mu$ are generators of the
Clifford algebra $Cl(1,3)$ satisfying
$\gam_\mu \gam_\nu + \gam_\nu \gam_\mu = 2 \eta_{\mu \nu}$.
An object $x^\mu \gam_\mu$ is a vector in $M_{1,3}$.

Instead of the basis vectors $\gam_\mu$, $\mu=0,1,2,3$, we can introduce
another set of basis vectors (forming the so-called Witt basis),
\be
    \nth_1 = \frac{1}{2}(\gam_0 + \gam_3), ~~~~~
    \nth_2 = \frac{1}{2}(\gam_1 + i \gam_2),
\lbl{2.1}
\ee
 \be
    {\bar \nth}_1 = \frac{1}{2}(\gam_0 - \gam_3), ~~~~~
    {\bar \nth}_2 = \frac{1}{2}(\gam_1 - i \gam_2),
\lbl{2.2}
\ee
which satisfy
\be
   \lbrace {\nth}_a,{\bth}_b \rbrace = \eta_{a b},~~~~
   \lbrace \nth_a, \nth_b \rbrace = 0,~~~~
   \lbrace \bth_a, \bth_b \rbrace = 0,
\lbl{2.3}
\ee
where $\eta_{a b} = {\rm diag} (1,-1),~~a,b=1,2$. Relations (\ref{2.3})
are fermionic anticommutation relations. 

Minkowski space $M_{1,3} \equiv M$ has been decomposed according to
$M = N {\dot +} P$ into two singular subspaces with null quadratic
forms.

We now observe that the product
\be
   f = \bth_1 \bth_2
\lbl{2.4}
\ee
satisfies
\be
     \bth_a f = 0,~~~~~a=1,2.
\lbl{2.5}
\ee
Therefore, $f$ can be interpreted as `vacuum', and $\bth_a$ can be interpreted as operators that annihilate $f$.   
   An object constructed as a superposition,
\be
    \Psi = (\psi^0 {\ul 1} + \psi^1 \nth_1 + \psi^2 \nth_2
    + \psi^{12} \nth_1 \nth_2) f,
\lbl{2.6}
\ee
is an element of a minimal left ideal of complexified $Cl(1,3)$;
it is a {\it spinor}. The notation of the components in Eq.\,(\ref{2.6})
is adapted to the fact that $\Psi$ is given in terms of a Clifford
algebra of a 2-dimensional vector space spanned by $\nth_1,~\nth_2$
acting on $f$. Because $\nth_1$ and $\nth_2$ are null vectors,
satisfying $\nth_1^2 = 0,~\nth_2^2 = 0$, the latter Clifford algebra is in
fact the Grassmann algebra $\bigwedge N$.

Because Eq.\,(\ref{2.6}) describes a 4-component spinor, it is convenient
to change the notation according to
\be
    \Psi = (\psi^1 {\ul 1} + \psi^2 \nth_1 \nth_2 + 
    \psi^3 \nth_1 + \psi^4 \nth_2 )f.
\lbl{2.7}
\ee
The even part of $\Psi$ is a left-handed spinor, whilst the odd part
is a right-handed spinor:
\be
    \Psi_L = (\psi^1 {\ul 1} + \psi^2 \nth_1 \nth_2) \bth_1 \bth_2,
\lbl{2.7a}
\ee
\be
    \Psi_R =  (\psi^3 \nth_1 + \psi^4 \nth_2 ) \bth_1 \bth_2.
\lbl{2.7b}
\ee
Defining $\gam_5 = \gam_0 \gam_1 \gam_2 \gam_3$, we have
\be
      i \gam_5 \Psi_L = - \Psi_L,
\lbl{2.7c}
\ee
\be
      i \gam_5 \Psi_R =  \Psi_R.
\lbl{2.7d}
\ee

To illustrate that $\Psi$ indeed behaves as a spinor under
rotation, let us consider as an example
\be
    R_{\gam_1 \gam_2} = {\rm e}^{\frac{1}{2} \gam_1 \gam_2 \varphi} =
    {\rm cos} \, \frac{\varphi}{2} + 
    \gam_1 \gam_2 {\rm sin} \,\frac{\varphi}{2},
\lbl{2.8}
\ee
which (after using $\gam_1 = \nth_2 +\bth_2,~~\gam_2 =(-i)(\nth_2 - \bth_2)$
that follow from inverting Eqs.\,(\ref{2.1}), (\ref{2.2})) becomes
\be
   R_{\gam_1 \gam_2} = {\rm e}^{\frac{i}{2} [\nth_2,\bth_2] \varphi } =
   {\rm cos} \, \frac{\varphi}{2} + 
     [\nth_2,\bth_2]{\rm sin} \,\frac{\varphi}{2}.
\lbl{2.9}
\ee
Then, we find
\be
    R_{\gam_1 \gam_2} \Psi = \left ( {\rm e}^{\frac{i \varphi}{2}} \psi^1
    {\ul 1} + {\rm e}^{-\frac{i \varphi}{2}} \psi^2 \nth_1 \nth_2 +
    {\rm e}^{\frac{i \varphi}{2}} \psi^3 \nth_2 +
    {\rm e}^{-\frac{i \varphi}{2}} \psi^4 \nth_2 \right ) f,
\lbl{2.10}
\ee
where we have taken into account $\nth_a f = 0$. In Eq.\,(\ref{2.10}), we
recognize the well-known transformation of a 4-component spinor.

\section{Four independent spinors}

Thus far, we have considered one possible decomposition of $M_{1,3}$ into
two singular subspaces, $N$ and $P$, that are spanned, respectively, over the
basis $(\nth_1,\nth_2)$ and $(\bth_1,\bth_2)$. There is another possibility,
namely, $(\nth_1,\bth_2)$ and $(\bth_1,\nth_2)$. We can thus consider
four different vacua,
\be
   f_1 = \bth_1 \bth_2\, ,~~~~~f_2 = \nth_1 \nth_2\, , ~~~~~ 
   f_3 = \nth_1 \bth_2\, , ~~~~~f_4 = \bth_1 \nth_2\, ,
\lbl{2.11}
\ee
and can construct four different kinds of spinors, each being in a different
minimal left ideal of $Cl(1,3)$ (more precisely, of its complexified version):
\be
    \Psi^1 = (\psi^{1 1}{\ul 1} + \psi^{2 1} \nth_1 \nth_2 + 
    \psi^{3 1} \nth_1 + \psi^{4 1} \nth_2 )f_1,
\lbl{2.12}
\ee
 \be
    \Psi^2 = (\psi^{1 2}{\ul 1} + \psi^{2 2} \bth_1 \bth_2 + 
    \psi^{3 2} \bth_1 + \psi^{4 2} \bth_2 )f_2,
\lbl{2.13}
\ee
\be
    \Psi^3 = (\psi^{1 3} \bth_1 + \psi^{2 3} \nth_2
             + \psi^{3 3} {\ul 1} + \psi^{4 3} \bth_1 \nth_2 ) f_3 ,
\lbl{2.14}
\ee
\be
    \Psi^4 = (\psi^{1 4} \nth_1 + \psi^{2 4} \bth_2
             + \psi^{3 4} {\ul 1} + \psi^{4 4} \nth_1 \bth_2 ) f_4 .
\lbl{2.15}
\ee
An arbitrary element of $Cl(1,3)$
is the sum of those independent spinors:
\be
   \Phi = \Psi^1 + \Psi^2 + \Psi^3 + \Psi^4 =
   \psi^{\alpha i} \xi_{\alpha i} \equiv \psi^{\tl A} \xi_{\tl A} ,
\lbl{2.16}
\ee
where
the set of elements $\lbrace {\ul 1} f_1, \nth_1 \nth_2 f_1, ...,
\nth_1 f_4, \bth_2 f_4, {\ul 1} f_4, \bth_1 \nth_2 f_4 \rbrace
\equiv \xi_{\tl A} \equiv \xi_{\alpha i}$, $~\alpha,~i = 1,2,3,4$,
forms a spinor basis of $Cl(1,3)$. We will call the object $\Phi$
a {\it generalized spinor}.

Spinors are thus just particular Clifford numbers. Usually a
Clifford number is given in terms of 16 basis elements
$\Gam_M \equiv ({\ul 1},\gam_\mu, \gam_{\mu \nu}, \gam_5 \gam_\mu,
\gam_5)$, but it can also be given in terms of the 16
basis elements $\xi_{\tl A} \equiv \xi_{\alpha i}$, where the first
index denotes four spinor components, whereas the second
index denotes four independent left minimal ideals of
$Cl(1,3)$. In general, $\psi^{\alpha i}$ are complex-valued spacetime
fields.

\section{Behavior of spinors under discrete Lorentz transformations}

We will now explore how the left-handed and right-handed spinors
(\ref{2.7a}),(\ref{2.7b}) transform under space inversion P, under time
reversal T, and under PT.

{\it Space inversion}, $\gam_0 \rightarrow \gam_0$, 
$\gam_r \rightarrow - \gam_r$, $r=1,2,3$
\bear
    &&\nth_1 \rightarrow \frac{1}{2} (\gam_0 - \gam_3) = \bth_1, \nonumber \\
    && \nth_2 \rightarrow \frac{1}{2} (-\gam_1 - i \gam_2) = - \nth_2 ,
    \nonumber \\
    &&\bth_1 \rightarrow \frac{1}{2} (\gam_0 + \gam_3) = \nth_1, \nonumber \\
    && \bth_2 \rightarrow \frac{1}{2} (-\gam_1 + i \gam_2) = - \bth_2 .
\lbl{4.1}
\ear
Using Eqs.\,(\ref{2.7a}),(\ref{2.7b}) and the above transformations
of the Witt basis, we find
\be
   \Psi_L \rightarrow \Psi'_L = 
   (- \psi^1 {\ul 1} + \psi^2 \bth_1 \nth_2) \nth_1 \bth_2 \equiv 
   {\Psi}_R^\natural \neq \Psi_R,
\lbl{4.2}
\ee
\be
    \Psi_R \rightarrow \Psi'_R = 
    (-\psi^3 \bth_1 + \psi^4 \nth_2) \nth_1 \bth_2 
    \equiv {\Psi}_L^\natural   \neq \Psi_L .
\lbl{4.3}
\ee

We see that the left-handed spinor $\Psi_L$ does not transform into the
right-handed $\Psi_R$ of Eq.\,(\ref{2.7b}). More precisely, it does not
transform into the right-handed spinor of the same ideal of $Cl(1,3)$.
Instead, it transforms into the spinor ${\Psi}_R^\natural$, which, according
to Eq.\,(\ref{2.14}), belongs to the third ideal. The transformed
spinor ${\Psi}_R^\natural$ is right-handed because it satisfies
\be
  i \gam_5 {\Psi}_R^\nl = 
  i \gam_5(- \psi^1 {\ul 1} + \psi^2 \bth_1 \nth_2) \nth_1 \bth_2 = 
  {\Psi}_R^\nl,
\lbl{4.4}
\ee
where we have used Eqs.\,(\ref{2.1}),(\ref{2.2}) and
$\lbrace \gam_5, \gam_\mu \rbrace = 0$, from which it follows that
$i \gam_5 \nth_1 \bth_2 = \nth_1 \bth_2$.

Analogously, the right-handed spinor $\Psi_R$ transforms into the
spinor ${\Psi}_L^\nl$, which, according to Eq.\,(\ref{2.14}),
also belongs to the third ideal and satisfies
\be
  i \gam_5 {\Psi}_L^\natural = i \gam_5 (-\psi^3 \bth_1+\psi^4\nth_2) 
  \nth_1 \bth_2   = -{\Psi}_L^\nl .
\lbl{4.5}
\ee
Therefore, ${\Psi}_L^\nl$ is a left-handed spinor.

By including a new index $j$ that runs over ideals, we have
\be
     \Psi^{j} = \Psi_L^j + \Psi_R^j\, , ~~~~ j =1,2,3,4.
\lbl{4.6}
\ee
Under the space inversion, $\Psi_L^j \rightarrow {\rm not}~ \Psi_R^j$,
and  $\Psi_R^j \rightarrow {\rm not} ~ \Psi_L^j$, but
$\Psi_L^j \rightarrow \Psi_R^{j'}$, and $\Psi_R^j \rightarrow 
\Psi_L^{j'}$,
where $j = 1,2,3,4$ and $j'=3,4,1,2$. Thus, a left (right)-handed spinor of
an ideal $j$ transforms into a right (left)-handed spinor of a different
ideal $j'$.

{\it Time reversal}, $\gam_0 \rightarrow  -\gam_0,~~\gam_r \rightarrow
 \gam_r,~~r=1,2,3$.
\bear
    &&\nth_1 \rightarrow \frac{1}{2} (-\gam_0 + \gam_3) = -\bth_1, \nonumber \\
    && \nth_2 \rightarrow \frac{1}{2} (\gam_1 + i \gam_2) = \nth_2 ,
    \nonumber \\
    &&\bth_1 \rightarrow \frac{1}{2} (-\gam_0 - \gam_3) = -\nth_1, \nonumber \\
    && \bth_2 \rightarrow \frac{1}{2} (\gam_1 - i \gam_2) = \bth_2 .
\lbl{4.7}
\ear 
By inserting the above transformations into Eqs.\,(\ref{2.7a}),(\ref{2.7b}),
we obtain
\be
   \Psi_L \rightarrow \Psi''_L = 
   (- \psi^1 {\ul 1} + \psi^2 \bth_1 \nth_2) \nth_1 \bth_2 = 
   {\Psi}_R^\sharp,
\lbl{4.8}
\ee
\be
    \Psi_R \rightarrow \Psi''_R = 
    (\psi^3 \bth_1 - \psi^4\nth_2) \nth_1 \bth_2 
    = - {\Psi}_L^\sharp .
\lbl{4.9}
\ee
Again, a left (right)-handed spinor transforms into a right (left)-handed spinor
of a different ideal. In the case illustrated above, a left (right)-handed
spinor of the first ideal transforms into a right (left)-handed spinor
of the third ideal. In general, under time reversal,
$\Psi_L^j \rightarrow \Psi_R^{j'}$ and $\Psi_R^j \rightarrow 
\Psi_L^{j'}$,
where $j = 1,2,3,4$ and $j'=3,4,1,2$.

{\it Space inversion and time reversal}, $\gam_0 \rightarrow - \gam_0,
~~\gam_r \rightarrow - \gam_r,~~r=1,2,3$.
\be
   \nth_a \rightarrow -\nth_a,~~~~  \bth_a \rightarrow -\bth_a,~~~a =1,2.
\lbl{4.10}
\ee
Then, we have $\Psi_L \rightarrow \Psi_L$ and $\Psi_R \rightarrow - \Psi_R$.

In general, under a Lorentz transformation,
including a discrete (i.e., improper) one,
an arbitrary element of $Cl(1,3)$, Eq.\,(\ref{2.16}), transforms
according to
\be
   \Phi = \psi^{\tl A} \xi_{\tl A} \rightarrow
   \Phi' = \psi^{\tl A} \xi'_{\tl A} = 
   \psi^{\tl A} {L_{\tl A}}^{\tl B} \xi_{\tl B} = \psi'^{\tl B} \xi_{\tl B},
\lbl{4.12}
\ee
where
\be
    \xi'_{\tl A} = {L_{\tl A}}^{\tl B} \xi_{\tl B},
\lbl{4.13}
\ee
\be
    \psi'^{\tl B} = \psi^{\tl A} {L_{\tl A}}^{\tl B}.
\lbl{4.14}
\ee
Here, ${\tl A} = \alpha i$ and ${\tl B} = \beta j$. 
Because ${\tl A}$ denotes the double index $\alpha i$, we can arrange
the basis elements $\xi_{\tl A} \equiv \xi_{\alpha i}$ into a matrix as
\be
  \xi_{\tl A} \equiv \xi_{\alpha i} = \begin{pmatrix}
   f_1    &      f_2   &      \bth_1 f_3      &     \nth_1 f_4 \\
 \nth_1 \nth_2 f_1 & \bth_1 \bth_2 f_2 &  \nth_2 f_3 &       \bth_2 f_4 \\
  \nth_1 f_1 &   \bth_1 f_2   &        f_3        &        f_4 \\
  \nth_2 f_1 &   \bth_2 f_2 & \bth_1 \nth_2 f_3  & \nth_1 \bth_2 f_4
  \end{pmatrix}
\lbl{4.15}
\ee
and components $\psi^{\tl A} \equiv \psi^{\alpha i}$ into a matrix as
\be
    \psi^{\alpha i} = \begin{pmatrix}
                       \psi^{11} & \psi^{12} & \psi^{13} & \psi^{14} \\
                       \psi^{21} & \psi^{22} & \psi^{23} & \psi^{24} \\
                       \psi^{31} & \psi^{32} & \psi^{33} & \psi^{34} \\
                       \psi^{41} & \psi^{42} & \psi^{43} & \psi^{44}
                       \end{pmatrix}.
\lbl{4.16}
\ee

For instance, under space inversion (\ref{4.1}), the Witt basis elements
transform according to (\ref{4.1}), and the basis elements
(\ref{4.15}) transform according to (2.13) into new basis elements,
\be
  \xi'_{\alpha i} =    
  \begin{pmatrix}
  -f_3     &    -f_4     &     -\nth_1 f_1   &    -\bth_1 f_2 \\
 \bth _1 \nth_2 f_3  & \nth_1 \bth_2 f_4  & \nth_2 f_1   &   \bth_2 f_2 \\
  -\bth_1 f_3   &    -\nth_1 f_4     &    -f_1     &        -f_2 \\
   \nth_2 f_3     &   \bth_2 f_4 & \nth_1 \nth_2 f_1 & \bth_1 \bth_2 f_2 
  \end{pmatrix},
\lbl{4.17}
\ee
whereas the components transform according to (\ref{4.14}) as
\be
   \psi^{\alpha i} =  \begin{pmatrix}
   -\psi^{33} & -\psi^{34} & -\psi^{31} & -\psi^{32} \\
   \psi^{43} & \psi^{44} & \psi^{41} & \psi^{42} \\
   -\psi^{13} & -\psi^{14} & -\psi^{11} & -\psi^{12} \\
     \psi^{23} & \psi^{24} & \psi^{21} & \psi^{22}
    \end{pmatrix}.
\lbl{4.18}
\ee

Because we consider the active transformations, then, according to (\ref{4.12}),
either Eq.\,(\ref{4.13}) or Eq.\,(\ref{4.14}) holds, but both cannot hold at
once. A generic Clifford number $\Phi$ of (\ref{2.16}), a sum of the
spinors of four different left ideals, thus transforms into a different
object $\Phi'$ that is expanded, according to Eq.\,(\ref{4.12}), either in terms of a
new basis (\ref{4.17}) and old components or in terms of the old basis
and new components (\ref{4.18}).

\section{An action for the generalized spinor field}

A possible action for a spacetime-dependent field $\psi^{\tl A}$
is\footnote{This action can be embedded into an even more general
action that depends on position in Clifford space, a manifold whose
tangent space at any point is a Clifford
algebra \ci{PavsicKaluza,PavsicKaluzaLong}.}
\be
    I[\psi_{\tl A}^*,\psi^{\tl B}] = \int \dd^4 x \, \sqrt{-g}\, i \,
     \psi_{\tl A}^* {(\gam^\mu)^{\tl A}}_{\tl B}\, \DD_\mu \psi^{\tl B}.
\lbl{5.1}
\ee
Here\,\ci{PavsicKaluza,PavsicKaluzaLong},
\be
    {(\gam^\mu)^{\tl A}}_{\tl B} = \langle {\xi^{\tl A}}^{\ddg} \gam^\mu
    \xi_{\tl B} \rangle_S = {\delta^i}_j {{(\gam^\mu )}^\alpha}_\beta \, ~~~
  \alpha , \beta , \gam, \delta = 1,2,3,4
\lbl{5.2}
\ee
are matrix elements of the 16-dimensional, reducible, representation
of the generators $\gam^\mu$. The operation $\langle A \rangle_S$ takes
the scalar part of a generic Clifford number $A$ and multiplies
the result by the dimension $n=4$ of the minimal left ideal
(i.e., the dimension of a spinor of $Cl(1,3)$). The symbol `$\ddg$'
denotes reversion, i.e., the operation that reverses the order
of all 1-vectors in an expression.

Indices are lowered and raised by the metric
$Z_{{\tl A}{\tl B}} = \langle \xi_{\tl A}^\ddg \xi_{\tl B} \rangle_S
= z_{ij} z_{\alpha \beta}$ and its inverse $Z^{{\tl A}{\tl B}}$.

The part of action (\ref{5.1}) that belongs to one
particular left ideal is obtained by fixing the ideal index,
e.g., $i=j=1$ in Eq.(\ref{5.2}), and writing $\psi_{\alpha 1}^*\equiv 
\psi_\alpha$, $\psi^{\beta 1}\equiv \psi^\beta$:
\be
    I[\psi_\alpha^*,\psi^\beta] = \int \dd^4 x \, \sqrt{-g} \, i \,
    \psi_\alpha^* {{(\gam^\mu )}^\alpha}_\beta \, \DD_\mu \psi^\beta.
\lbl{5.3}
\ee
Here, ${{(\gam^\mu )}^\alpha}_\beta$ are the ordinary Dirac matrices satisfying
${{(\gam^\mu )}^\alpha}_\gam {{(\gam^\nu )}^\gam}_\beta$
$+ {{(\gam^\nu )}^\alpha}_\gam {{(\gam^\mu )}^\gam}_\beta =
2 \eta^{\mu \nu} {\delta^\alpha}_\beta $. A complex conjugate spinor
with the lower index $\alpha$ can be expressed in terms of a spinor
with the upper index $\alpha$ as $\psi_\alpha^* 
 = z_{\alpha \beta} {\psi^*}^\beta$.
Because, in particular classes of representations,
$z_{\alpha \beta}$ equal ${{(\gam^0 )}^\alpha}_\beta$, we have
that $\psi_\alpha^*$ corresponds to $\bar \psi = \psi^\dg \gam^0$.

The covariant derivative in (\ref{5.1}) reads
\be
    \DD_\mu \psi^{\tl A} = \p_\mu \psi^{\tl A} + {{G_\mu}^{\tl A}}_{\tl B} 
     \psi^{\tl B},
\lbl{5.4}
\ee
where ${{G_\mu}^{\tl A}}_{\tl B} \equiv {{G_\mu}^{\alpha i}}_{\beta j}$
is a gauge field that, in general, couples to the spinors of different
left ideals (denoted by indices $i$ or $j$).

The interactive term Lagrangian in the action (\ref{5.1}) can be
written explicitly as
\be
    {\cal L}_{\rm int} = \psi_{\alpha i}^* {(\gam^\mu)^{\alpha i}}_{\beta j}
     {{G_\mu}^{\beta j}}_{\rho k} \psi^{\rho k} = \psi_{\alpha i}^*
    {(\gam^\mu)^{\alpha}}_{\beta} {{G_\mu}^{\beta i}}_{\rho k} \psi^{\rho k}.
\lbl{5.5}
\ee
If acting within the same left ideal, the field $ {{G_\mu}^{\beta i}}_{\rho k}$
for $i=k$ behaves as the ordinary spin connection in curved
spacetime. Otherwise, it behaves as a gauge field due to the
``internal" local gauge transformations\,\ci{PavsicKaluza,PavsicKaluzaLong} that
transform one ideal into the other. The group of the latter transformations
contain SU(2) as a subgroup.

Let us write Eq.\,(\ref{5.5}) in a more compact form,
\be
   {\cal L}_{\rm int} = {\bar \psi}_i \, \gam^\mu \, {{G_\mu}^i}_k \, \psi^k,
\lbl{5.6}
\ee
where all quantities now contain the implicit spinor indices $\alpha$,$\beta$.
In the chiral representation,
\be
   \gam^\mu =  \begin{pmatrix}
    0      &     \sigma^\mu \\
   {\bar \sigma}^\mu & 0
   \end{pmatrix}
    ~,~~~~\sigma^\mu = (1,\sigma^r)~,
       ~~~{\bar \sigma}^\mu = (1,-\sigma^r)~,~~~r=1,2,3,
\lbl{5.7}
\ee
where $\sigma^r$ are the Pauli $2 \times 2$ matrices. If we write the
gauge fields in the following block matrix form,
\be
    {{G_\mu}^i}_k =  
    \begin{pmatrix}
     {{a_\mu}^i}_k  &   {{b_\mu}^i}_k \\
    {{c_\mu}^i}_k   &  {{d_\mu}^i}_k  
    \end{pmatrix},
\lbl{5.8}
\ee
then
\be
    \gam^\mu \, {{G_\mu}^i}_k \psi^k = 
       \begin{pmatrix} 
       \sigma^\mu {{c_\mu}^i}_k      &     \sigma^\mu {{d_\mu}^i}_k \\
       {\bar \sigma}^\mu {{a_\mu}^i}_k  &  {\bar \sigma}^\mu {{b_\mu}^i}_k 
       \end{pmatrix}
       \begin{pmatrix}
    \chi_L^k \\
    \chi_R^k
    \end{pmatrix}.
\lbl{5.9}
\ee
Here, ${{a_\mu}^i}_k$, ${{b_\mu}^i}_k$, ${{c_\mu}^i}_k$, ${{d_\mu}^i}_k$
are $2\times 2$ matrices, whilst $\chi_L^k$, $\chi_R^k$ are 2-component
Weyl spinors. Because ${\bar \psi}_i = (\chi_{i R}^\dg , \chi_{i L}^\dg )$,
we have
\be
    {\cal L}_{\rm int} = 
    \chi_{i R}^\dg \, \sigma^\mu \, {{c_\mu}^i}_k \chi_L^k +
    \chi_{i R}^\dg \, \sigma^\mu \, {{d_\mu}^i}_k \chi_R^k +
     \chi_{i L}^\dg \, {\bar \sigma}^\mu \, {{a_\mu}^i}_k \chi_L^k +
     \chi_{i L}^\dg \, {\bar \sigma}^\mu \, {{b_\mu}^i}_k \chi_R^k.
\lbl{5.10}
\ee
This interactive Lagrangian has an interesting structure that involves
spinor degrees of freedom and ``internal" degrees of freedom
associated with different left ideals (denoted by indices $i,~k$) and
various gauge fields. Left-handed or right-handed spinors of one
left ideal, denoted by $k$, are coupled (via various gauge fields)
to the corresponding spinors
of another left ideal, denoted by $i$ (where, in particular, it can
be $i=k$).

If we split the indices according to $i={\bar i},{\ul i}$
and $k = {\bar k},{\ul k}$, where
${\bar i},{\bar k}=1,2$ and ${\ul i}, {\ul k}=3,4$, then the Lagrangian
(\ref{5.10}) can be written in the form of
\bear
    {\cal L}_{\rm int} = &&
    \chi_{{\bar i} R}^\dg \, \sigma^\mu \, {{c_\mu}^{\bar i}}_{\bar k} 
    \chi_L^{\bar k} +
    \chi_{{\bar i} L}^\dg \, {\bar \sigma}^\mu \, 
    {{b_\mu}^{\bar i}}_{\bar k} \chi_R^{\bar k} +
    \chi_{{\ul i} R}^\dg \, \sigma^\mu \, {{c_\mu}^{\ul i}}_{\ul k} 
    \chi_L^{\ul k} +
    \chi_{{\ul i} L}^\dg \, {\bar \sigma}^\mu \, 
    {{b_\mu}^{\ul i}}_{\ul k} \chi_R^{\ul k} \nonumber \\
   &+& \chi_{{\bar i} L}^\dg \, {\bar \sigma}^\mu \, {{a_\mu}^{\bar i}}_{\bar k} 
    \chi_L^{\bar k} +
    \chi_{{\bar i} R}^\dg \, \sigma^\mu \, 
    {{d_\mu}^{\bar i}}_{\bar k} \chi_R^{\bar k} +
    \chi_{{\ul i} L}^\dg \, {\bar \sigma}^\mu \, {{a_\mu}^{\ul i}}_{\ul k} 
    \chi_L^{\ul k} +
    \chi_{{\ul i} R}^\dg \, \bar \sigma^\mu \, 
    {{d_\mu}^{\ul i}}_{\ul k} \chi_R^{\ul k}.    
\lbl{5.11}
\ear
The bared indices denote the 1st and 2nd ideals, whereas the
underlined indices denote the 3rd and 4th ideals.

One can show that a rotation from the
right, $\Phi \rightarrow \Phi' = \Phi R_{\gam_r \gam_s}$, in general
transforms the 1st into the 2nd and the 3rd into the 4th ideal. The left-handed
field $\chi_L^{\bar k}$, ${\bar k}=1,2$ forms the fundamental
representation of SU(2), and it is coupled via the SU(2) gauge
field ${{a_\mu}^{\bar i}}_{\bar k}$ to $\chi_{{\bar i} L}^\dg$
and via ${{c_\mu}^{\bar i}}_{\bar k}$ to $\chi_{{\bar i} R}^\dg$.
The analogous holds for the fields $\chi_L^{\ul k}$, $\chi_R^{\bar k}$,
and $\chi_R^{\ul k}$.

We have seen in Eqs.\,(\ref{4.17}),(\ref{4.18}) that space inversion
transforms the 1st ideal into the 3rd one and the 2nd ideal into the
4th one. Besides that, space inversion exchanges the upper half of the
4-component spinor with the lower half, i.e., it exchanges
$\chi_L^k \rightarrow \chi_R^{k'}$ and $\chi_L^k \leftarrow \chi_R^{k'}$.
Instead of the inversion (\ref{4.1}), it is convenient to apply the
reflection $\gam_0 \rightarrow \gam_0$, $\gam_1 \rightarrow \gam_1$,
$\gam_2 \rightarrow \gam_2$, $\gam_3 \rightarrow -\gam_3$, because
then no changes of sign take place in the transformations of
$\xi_{\alpha i}$, $\psi^{\alpha i}$, and ${G^{\alpha i}}_{\beta j}$.
Using (\ref{4.13}), one can show that
\be
    {{G_\mu}^{\alpha i}}_{\beta j} \rightarrow {{G'_\mu}^{\alpha i}}_{\beta j}
    = {L_{\delta k}}^{\alpha i} {L^{\gam \ell}}_{\beta j}
    {{G_\mu}^{\delta k}}_{\gam \ell}.
\lbl{5.12}
\ee
So, we find that 
\be
   {{a_\mu}^{\bar i}}_{\bar k} \rightarrow {{d_\mu}^{\ul i}}_{\ul k}\;,~~~~
      {{a_\mu}^{\ul i}}_{\ul k} \rightarrow {{d_\mu}^{\bar i}}_{\bar k},
\lbl{5.13}
\ee
\be
  {{c_\mu}^{\bar i}}_{\bar k} \rightarrow {{b_\mu}^{\ul i}}_{\ul k}\;,~~~~
      {{b_\mu}^{\ul i}}_{\ul k} \rightarrow {{c_\mu}^{\bar i}}_{\bar k}.
\lbl{5.14}
\ee

The interaction Lagrangian (\ref{5.10}) (which can be written in the form of
(\ref{5.11})) is thus invariant under space inversion. A notorious
feature of this Lagrangian is that an ordinary right-handed field
$\chi_R^{\bar k}$, ${\bar k} =1,2$, is not coupled to the same gauge
field as the left-handed field $\chi_L^{\bar k}$. The right-handed field
$\chi_R^{\bar k}$ behaves as a ``singlet" with respect to the SU(2)
gauge field ${{a_\mu}^{\bar i}}_{\bar k}$, which the left-handed field
$\chi_L^{\bar k}$ is coupled to. Instead, $\chi_R^{\bar k}$ behaves
as an SU(2) doublet with respect to another gauge field, namely,
${{d_\mu}^{\bar i}}_{\bar k}$.

Lagrangian (\ref{5.11}) contains, amongst others, a term
$\chi_{{\bar i} L}^\dg \, {\bar \sigma}^\mu \, {{a_\mu}^{\bar i}}_{\bar k} 
\chi_L^{\bar k}$. As already mentioned, ${{a_\mu}^{\bar i}}_{\bar k}$
is a $2 \times 2$ matrix in the spinor indices ${\bar \alpha},~{\bar \beta}
=1,2$. If, in particular, ${{a_\mu}^{\bar i}}_{\bar k}$ is a diagonal
matrix in indices ${\bar \alpha},~{\bar \beta}$, then the term
$\chi_{{\bar i} L}^\dg \, {\bar \sigma}^\mu \, {{a_\mu}^{\bar i}}_{\bar k} 
\chi_L^{\bar k}$ is just like the interactive Lagrangian for weak
interactions, ${{a_\mu}^{\bar i}}_{\bar k}$ being the weak SU(2)
gauge fields ${{W_\mu}^{\bar i}}_{\bar k}$.

However, Lagrangian (\ref{5.11}) contains other terms besides the one
describing the usual weak interaction. It is beyond the scope of this
Letter to investigate the physical meaning of those extra terms.
One possibility is that due to certain superselection rules, the fields
${{c_\mu}^i}_k$ and ${{b_\mu}^i}_k$ vanish. Another possibility is
that Eq.\,(\ref{5.11}) predicts the existence of new interactions.
From now on, we will only consider the part of ${\cal L}_{\rm int}$ that contains
the fields ${{a_\mu}^i}_k$ and ${{d_\mu}^i}_k$.

We adopt an interpretation that the group SU(2) of
$R_{\gam_r \gam_s} \in {\rm SU(2)}$ that acts on a generalized spinor
$\Phi$ from the right and that ``mixes" either indices ${\bar i},{\bar k}$
or ${\ul i},{\ul k}$ is the gauge group of weak interactions.
Because we know from experiments that right-handed ordinary fields do not interact
under the weak SU(2) gauge fields, we will assume
${{d_\mu}^{\bar i}}_{\bar k} =0$. In addition, let us also assume
${{a_\mu}^{\ul i}}_{\ul k} =0$. Then, the Lagrangian
(\ref{5.11}) contains the following part:
\be
   {\cal L}_{\rm int}^W =
   \chi_{{\bar i} L}^\dg \, {\bar \sigma}^\mu \, {{a_\mu}^{\bar i}}_{\bar k} 
    \chi_L^{\bar k} +
    \chi_{{\ul i} R}^\dg \,  \sigma^\mu \, {{d_\mu}^{\ul i}}_{\ul k} 
    \chi_R^{\ul k}
\;, ~~~~{\bar i},{\bar k}=1,2\;, ~~{\ul i},{\ul k}=3,4.
\lbl{5.15}
\ee
Here, we can adopt an interpretation that
$\chi_L^{\bar k}$ are spinor fields describing ordinary particles
coupled to the ordinary weak SU(2) gauge fields, whereas $\chi_R^{\ul k}$
are {\it mirror} fields describing mirror particles that are
coupled to the mirror gauge fields ${{d_\mu}^{\ul i}}_{\ul k}$.
Notice the difference between the barred and underlined indices.
In other words, spinor fields of the 1st and 2nd ideals describe ordinary
particles (and antiparticles), whereas spinor fields of the 3rd and 4th ideals
describe mirror particles (and mirror antiparticles).

Putting it into a more compact notation, we write (\ref{5.15}) as
\be
  {\cal L}_{\rm int}^W = \chi_L^\dg {\bar \sigma}^\mu W_\mu \chi_L
      + {\chi^\natural}_R^\dg \sigma^\mu \, W_\mu^\natural\, \chi_R^\natural,
\lbl{5.16}
\ee
where `$\natural$' denotes mirror fields, so that
$\chi_L \equiv \chi_L^{\bar k}$ and $\chi_R^\nl \equiv \chi_R^{\ul k}$.
 Here, $W_\mu$ and $W_\mu^\natural$
are SU(2) gauge fields, and thus they are $2 \times 2$ matrices in
indices ${\bar i},{\bar k}$ and ${\ul i},{\ul k}$, respectively, i.e.,
$W_\mu \equiv {{a_\mu}^{\bar i}}_{\bar k}$ and
$W_\mu^\natural \equiv {{d_\mu}^{\ul i}}_{\ul k}$.

The action (\ref{5.1}) then contains the following part:
\be
     I^W = \int \dd^4 x \, \sqrt{-g} \, i \, \left (
     {\bar \psi}_L \gam^\mu \p_\mu \psi_L +
     {\bar \psi}_L \gam^\mu\, W_\mu \psi_L +    
     {\bar \psi}_R^\natural \, \gam^\mu \p_\mu \psi_R^\natural +
     {\bar \psi}_R^\natural \, \gam^\mu W_{\mu}^\nl \psi_R^\natural  \right ), 
\lbl{5.17}
\ee
where
\bear
    &&\psi_L = \frac{1}{2} (1 - i \gam_5) \psi = 
    \begin{pmatrix}
     \chi_L \\
        0 
     \end{pmatrix}
      \equiv \begin{pmatrix}
      \chi_L^{\bar k} \\
        0  
        \end{pmatrix}
\lbl{5.18a}\\
      &&{\bar \psi}_L = (0, \chi_L^\dg)
         \equiv (0, {\chi_L^{\bar k}}^\dg)
 \lbl{5.18b}\\
      &&\psi_R^\natural = \frac{1}{2} (1 + i \gam_5) \psi^\natural
     =  \begin{pmatrix}
         0 \\
       \chi_R^\natural
       \end{pmatrix} \equiv
     \begin{pmatrix}
         0 \\
       \chi_R^{ {\ul k}}       
       \end{pmatrix}
\lbl{5.18c}\\
      &&{\bar \psi}_R^\natural = ({\chi_R^\natural}^\dg ,0) 
       \equiv ({\chi_R^{\ul k}}^\dg , 0) .
\lbl{5.18d}
\ear
In Eq.\,(\ref{5.16}), we identified the weak gauge field $W_\mu$ with
the diagonal part (with respect to hidden indices ${\bar \alpha}, {\bar \beta}$) 
of ${{a_\mu}^{\bar i}}_{\bar k}$ and the
mirror gauge field $W_\mu^{\nl}$ with the diagonal part of
${{d_\mu}^{\ul i}}_{\ul k}$. Remember that the latter gauge fields, in addition to being matrices
in the ``internal" indices ${\bar i},~{\bar k}$ and ${\ul i},~{\ul k}$,
are also matrices in the spinor indices ${\bar \alpha},~{\bar \beta}=1,2$
(that, for simplicity, are not explicitly displayed). On the other hand,
if we take $i=k$ and allow for ${\bar \alpha \neq \bar \beta}$, then
${{a_\mu}^i}_k$ and ${{d_\mu}^i}_k$ correspond to the ordinary
spin connections\footnote{This can be shown by observing
that for $i=k$, the generalized spin connection
${{G_\mu}^{\alpha i}}_{\beta k}$ is just the
ordinary spin connection ${{\Gam_\mu}^\alpha}_\beta =
\frac{1}{2}{\omega_\mu}^{a b} [\gam_a,\gam_b]$; then, by taking
into account Eqs.\,(\ref{5.7}), one finds that in Eq.\,(\ref{5.9}),
the part with ${{a_\mu}^i}_k$ and ${{d_\mu}^i}_k$, for $i=k$,
corresponds to the contribution of the ordinary spin connection.    }. 
    
The action (\ref{5.17}), which besides the ordinary particles described by
$\psi_L$ and the ordinary gauge fields $W_\mu$ contains the corresponding
mirror partners $\psi_R^\natural$ and $W_\mu^\natural$, encodes
the proposal by Kobzarev et al.\,\ci{Kobzarev}. Later, a model essentially
equivalent to that given by the action (\ref{5.17}) was proposed
by Foot et al.\,\ci{Foot}. In this paper, we have shown
how such a system with ordinary and mirror particles coupled to
the corresponding gauge fields can be described within a
framework based on the algebraic spinors and their behavior under
space inversion.

\section{Discussion and conclusion}

The observation that space inversion P and time reversal T transform
a spinor of one left ideal of $Cl(1,3)$ into a spinor of another left
ideal brings a novel insight on the invariance of nature under P and T
that, to my knowledge, has not yet been explicitly pointed out. The idea
that the invariance of weak interactions under P and T can be
restored by introducing mirror particles has been considered by Lee
and Yang\,\ci{LeeYang} and later in
Refs.\,\ci{Kobzarev}--\ci{Foot}.
In this letter, we provide a theoretical basis for such a model
that comes naturally from the well-known theory of algebraic
spinors, i.e., the objects embedded within the Clifford algebra
that operates on them.

Electron spin was first postulated based on experimental evidence
in atomic spectra and was later explained by the Dirac equation, which
provided a deep theoretical insight. We have a somewhat analogous
situation with mirror particles that were postulated to restore
left-right symmetry. In this paper, we have shown that the generalized
spinors, which encompass all four left ideals, describe not only
the ordinary spin and weak isospin but also
the ordinary and mirror particles. The corresponding
SU(2) gauge fields are incorporated in the generalized spin connection.
One further step that should be done is to include the U(1) gauge field
and to show precisely how the electric charge and the electromagnetic field
occur within this Clifford algebra-based theory. To include
the color SU(3) as well, an extension of $Cl(1,3)$ or its
complex version $Cl(4,\mathbb{C})$
to a suitable higher dimensional Clifford algebra, such as
$Cl(8,\mathbb{C})$, would be necessary.

\centerline{\bf Ackonwledgement}

This work was supported by the Ministry of High Education,
Science and Technology of Slovenia.

{\bf Note added}

In the published version of this paper (Phys.Lett. B 692 (2010) 212-217))
there is some explanation justifying the validity of Eq.\,(\ref{4.1}).
The new text introducing Section 4 is the following:

\begin{quote}
We will now explore how the left-handed and right-handed spinors
(\ref{2.7a}),(\ref{2.7b}) transform under space inversion P, under time
reversal T, and under PT. A generic transformation of the Lorentz group,
which include the latter improper transformation, transforms a vector
$a = a^\mu \gam_\mu$ into a new vector $a' =S a S^{-1} =
a^\mu S \gam_\mu S^{-1} = a^\mu {L^\nu}_\mu \gam_\nu$. We can rewrite this
in the form $a'= a'^\nu \gam_\nu$, where $a'^\nu ={L^\nu}_\mu a^\mu$, or
in the form $a' = a^\mu \gam'_\nu$, where $\gam'_\nu ={L^\nu}_\mu \gam_\nu$.
In the former case the vector components $a^\mu$ transform into new
components $a'^\mu$, the basis vectors $\gam_\mu$ remaining unchanged,
whereas in the latter case the components $a^\mu$ remain the same and the
basis vectors $\gam_\mu$ transform into new basis vectors $\gam'_\mu$.
In the following we will consider the latter case. 
\end{quote}

At the end of the published paper there is some discussion about how massless
fields in the action (\ref{5.1}) could become massive:

\begin{quote}
In order to obtain a realistic theory,
we should also investigate how fermionic fields, described by
the action (\ref{5.1}), acquire masses. According to the standard procedure
we have to include the Higgs fields into the action.
Another possibility is to generalize
the latter action to the 16-dimensional Clifford space;
the effective 4-dimensional mass then comes from the extra dimensions
of Clifford space\,\ci{PavsicKaluza}.
\end{quote}

\end{document}